# Revealing magnetic component in crystalline Fe-gluconate

S. M. Dubiel[1,2,*], J. Żukrowski[2] and Maria Bałanda[3]

[1]AGH University of Science and Technology, Faculty of Physics and Applied Computer Science, al. A. Mickiewicza 30, 30-059 Kraków, Poland, [2]Academic Centre of Materials and Nanotechnology, al. A. Mickiewicza 30, 30-059 Kraków, Poland, [3]Institute of Nuclear Physics, Polish Academy of Sciences, 31-342 Kraków, Poland

**Abstract**

Low temperature Mössbauer spectroscopic and magnetization measurements were performed on a crystalline sample of Fe-gluconate. Fe atoms were revealed to exist in two "phases" i.e. a major (~90-94%) and a minor (~6-10%). Based on values of spectral parameters the former can be regarded as ferrous and the latter as ferric ions. A sub spectrum associated with the ferric "phase" shows a significant broadening below ~30 K corresponding to ~7.5 kGs. A magnetic origin of the effect was confirmed by the magnetization measurements. Evidence on the effect of the magnetism on the lattice vibrations of Fe atoms in both "phases" was found. The Debye temperature, $T_D$, associated with the vibrations of $Fe^{2+}$ ions is by a factor of ~2 smaller in the temperature range below ~30 K than the one determined from the data measured above ~30 K. Interestingly, the $T_D$-value found for the $Fe^{3+}$ ions from the data recorded below ~30 K is about two times smaller than the corresponding value determined for the $Fe^{2+}$ ions.

\* Corresponding author: Stanislaw.Dubiel@fis.agh.edu.pl

# 1. Introduction

Ferrous gluconate (Fe-gluconate) is a salt of the gluconic acid. Its chemical formula reads as $C_{12}H_{22}FeO_{14} \cdot xH_2O$ with $0 \leq x \leq 2$. Its molar mass depends on the value of $x$ and varies between 446.14 and 482.19 g·mol$^{-1}$ for $x$=0 (dehydrated) and $x$=2 (fully hydrated), respectively. The compound has chiefly applications in medical and food additive industries. Regarding the former it has been satisfactorily used in the cure of hypochromic anemia and sold under various trade names e. g. Ascofer, Fergon, Ferate, Ferralet, FE-40, Gluconal FE and Simron, to list some of them. Respecting the latter, it has been applied for coloring foods, e. g. Black olives and beverages. Interestingly, Fe-gluconate was also used as an effective inhibitor for carbon steel [1], gluconate-based electrolytes were also successfully used to electroplate various metals [2] or alloys [3]. Iron, whose content lies between 11.8 and 12.5 percent, is present in two forms: a major ferrous $Fe^{2+}$ or Fe(II) ion and a minor ferric ($Fe^{3+}$) or Fe(III) ion. The relative contribution of the minor fraction amounts to ~10-15%, as detected by Mössbauer spectroscopy [4-8]. Its origin is unknown and it can either be soluble or insoluble. Following clinical studies, medicaments based on ferric iron have poorer absorption than those containing the ferrous ions [9]. Consequently, their effectiveness in the treatment of anemia diseases is less efficient. In other words, presence of the ferric ions in the Fe-gluconate is undesired from the medical viewpoint. In these circumstances any experiment aimed at the identification of the minor fraction is of importance as it can help to produce a ferric-free compound, or, at least, reduce its concentration. In a given structure of the Fe-gluconate, that can be either crystalline [10] or amorphous [11], the ferric ions should be stronger bonded than the ferrous ones, hence their lattice-dynamical properties and, in particular, a value of the Debye temperature, should be higher than that of the ferrous ions. The Mössbauer spectroscopy has been recognized as a relevant technique to investigate the issue. However, our recent study performed on a crystalline form of the Fe-gluconate in the temperature range of 80-310K did not show any measurable difference in the lattice-dynamical behavior of the two types of Fe-ions [12]. In order to shed more light on the issue we have performed similar measurements on the same sample of this compound but in a lower temperature range viz. 5-119 K. Results we have obtained are presented and discussed in this paper.

## 2. Experimantal

### 2.1. Sample

Fe-gluconate, courtesy of the Chemistry and Pharmacy Cooperative ESPEFA (Krakow, Poland), which uses it for a production of the iron supplement Ascofer®, was subject of the present study. X-ray diffraction pattern registered at room temperature (Fig. 1) on a powdered sample gave evidence that its structure was perfectly crystalline.

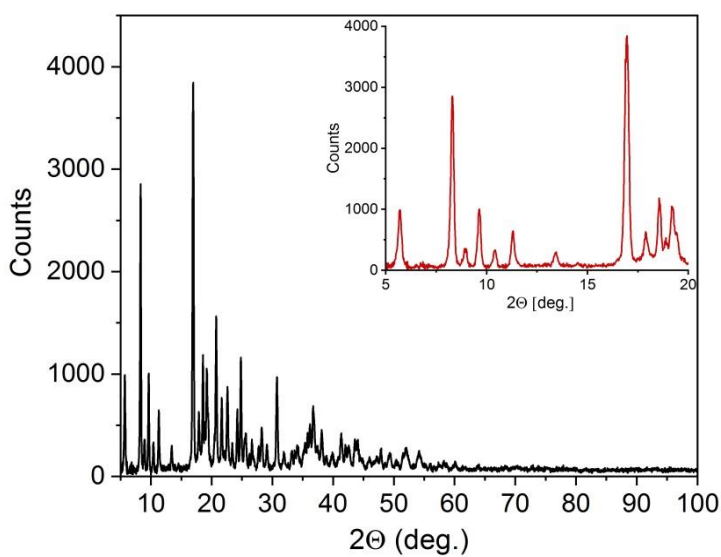

Fig. 1 X-ray diffraction pattern recorded at room temperature on a powdered sample of Fe-gluconate.

### 2.2. Mössbauer Spectra and Analysis

Mössbauer spectra, examples of which are shown in Figs. 2 and 3, were recorded in a transmission geometry by means of a standard (Wissel GmbH) spectrometer and a drive working in a sinusoidal mode. Each spectrum was recorded in a 1024 channels and 14.4 keV gamma rays were supplied by a $^{57}$Co/Rh source whose activity enabled recording a statistically good spectrum within a 1-2 days run. The spectra were measured in the temperature interval of 5-119 K on the sample placed in the Janis

Research 850-5 Mössbauer Refrigerator System. Temperature was stabilized to the accuracy of ±0.1 K.

The spectra were analyzed using a least-squares fitting procedure in two ways A and B. Concerning A, the spectra were fitted to three doublets, D1, D2 and D3. The first two were associated with the outermost lines constituting a major component of each spectrum and D3 with a minor component of which one line is visible between the two lines of the major component. As free parameters were treated: center shift (CS1, CS2, CS3), quadrupole splitting (QS1, QS2, QS3), line width (G1, G2, G3), relative contribution (A1, A2, A3). In the analysis B, which is based on an integral exchange method, the minor component, due to a significant broadening of the linewidth observed at lower temperatures (<~30 K), was fitted to a sextet. In this case as free parameters were treated: the main component of the electric field gradient ($V_{zz}(1)$, $V_{zz}(2)$, $V_{zz}$), line width (G1, G2, G), the center shift common to D1 and D2 (CS12), the center shift of the sextet (CS), the relative contribution of D1 and D2 common to D1 and D2 (A12) and the relative contribution of the sextet (A). Both procedures, from the statistical point of view, resulted in very good fits, yet the analysis in terms of the procedure B, reproduced better the visible line of the minor phase in the spectra recorded at low *T* i.e. where the broadening occurs – see Fig. 3.

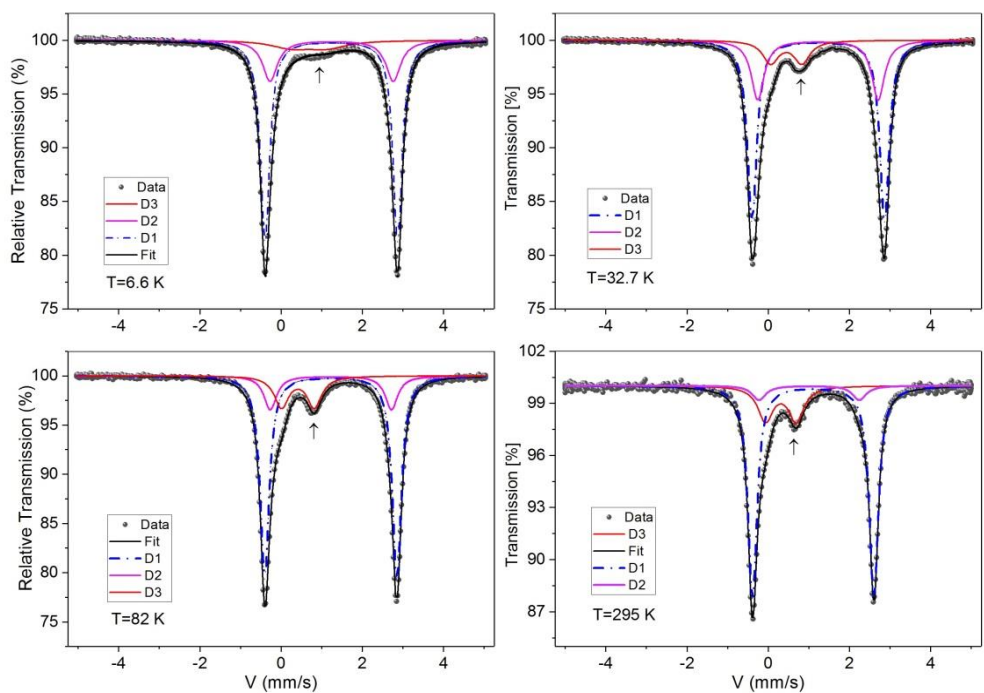

Fig. 2 $^{57}$Fe site Mössbauer spectra recorded at various temperatures on an untreated (as received) sample of a crystalline Fe-gluconate. The spectra were fitted to 3 doublets D1, D2 and D3. An arrow indicates a visible line of D3. Note a broadening of this line at 6.6 K.

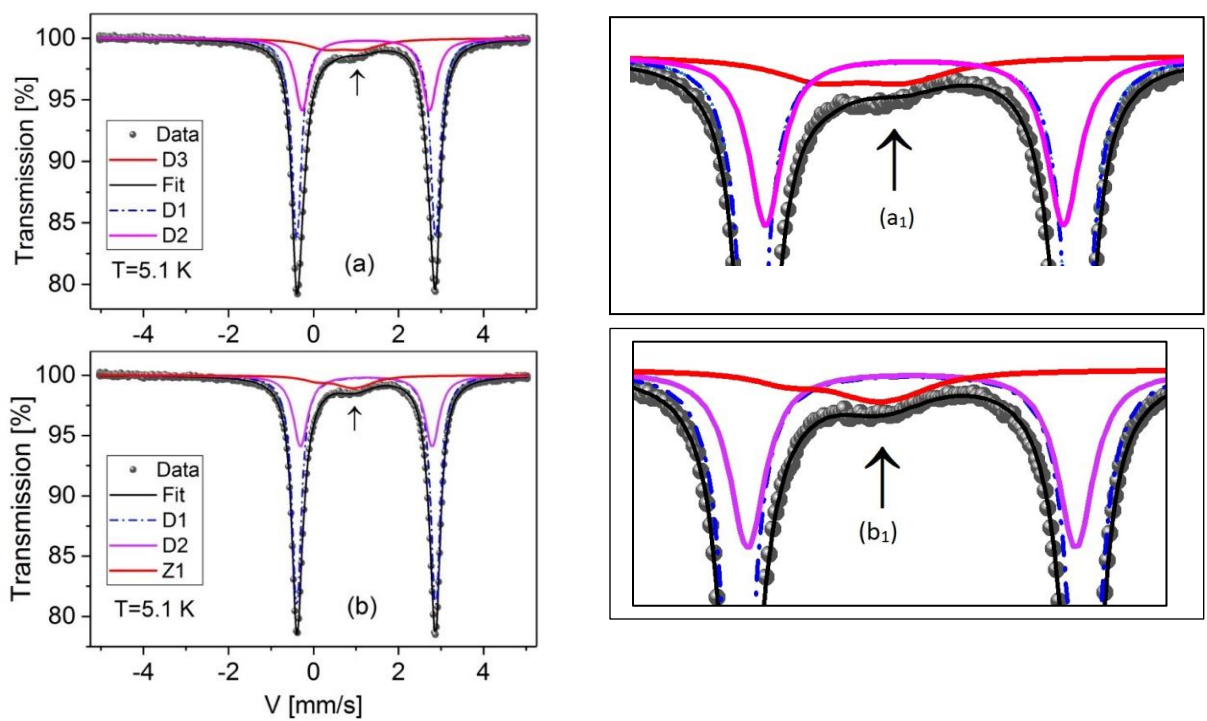

Fig. 3 $^{57}$Fe site Mössbauer spectra recorded at 5.1 K on the studied sample. Figure (a) shows an analysis to 3 doublets (D1, D2, D3) while (b) illustrates the analysis to two doublets (D1 and D2) and one sextet (Z1). Note that the quality of the fit is better in (b) – see the right hand panel.

Spectral parameters obtained with the two fitting procedures are displayed as plots and also in Table 1 (procedure B).

**2.3 Magnetic measurements**

The measurements were performed on a powder sample using a Quantum Design SQUID magnetometer. The temperature dependence of static magnetic susceptibility, $\chi_{DC}$ = M/H in zero-field cooled (ZFC) and field-cooled (FC) mode in the field H = 50 Oe and 100 Oe was measured in the *T*-range of 2-300K. The temperature independent contribution to the magnetic susceptibility of the samples was estimated and subtracted. AC susceptibility at the frequency 111 Hz and amplitude of the driving field 3 Oe was measured in the temperature range of 2-80 K. Isothermal magnetization curves, M(H), up to the field of 70 kOe were measured at 2 K, 5 K, 10 K, 20 K, and 40 K.

## 3. Results and Discussion

### 3.1. Spectral parameters

#### 3.1.1. Abundance

Temperature dependence of relative abundances of the three components as received from the analysis of the spectra using procedure A is displayed in Fig. 4a. We note that the contribution of D1, A1, lies between ~60% for T <~60 K, and ~70% for T > ~80 K. Similarly, the contribution of D2, A2, stays constant at the value of ~30 % up to ~60 K, and at the value of ~10% above ~80 K. The change of A1 and A2 in the temperature between ~60 and ~80 K is unknown. But as can be seen in Fig. 4a, the abundance of the minor component, A3, shows some anomaly at ~80 K and also it crosses with A2. Perhaps the anomalous behavior reflects some structural changes? Corresponding X-ray diffraction measurements are underway. Figure 4b illustrates a comparison between the abundance of the minor component as found with the two fitting procedures. One can easily notice that (a) for all temperatures the abundance found with procedure A is by ~5% higher than the one determined with procedure B, and (b) there is an anomaly below ~25 K viz. the abundance decreases. This means that in the temperature interval of 5-25 K, i.e. where a broadening of the line in the minor component is observed, the detectability of this component decreases. This in turn implies that the recoil-free factor decreases on lowering $T$. This effect, sometimes termed as a lattice softening, is known to occur on a transition from a paramagnetic to a magnetic state e. g. [13].

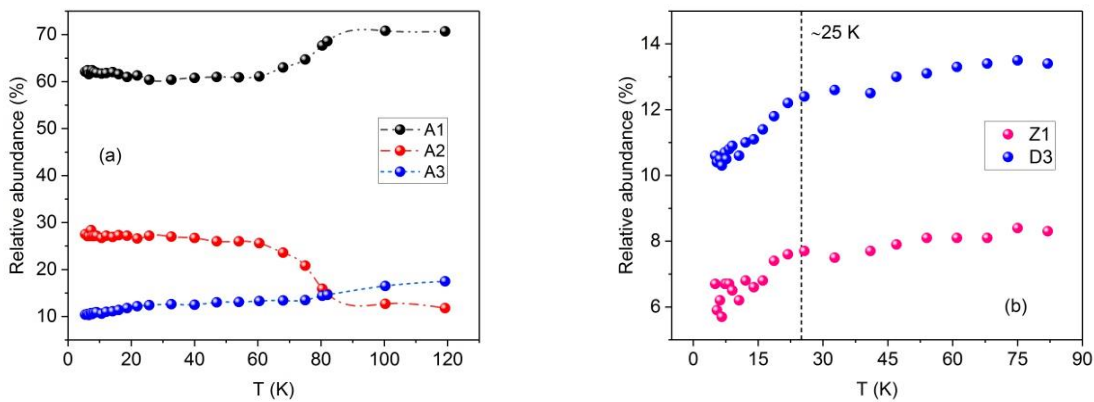

Fig. 4 (a) Relative abundances of the three components, A1, A2 and A3 as obtained with the fitting procedure A. Plot (b) shows a comparison between the abundance of the minor component as found with the procedure A, D3, and the procedure B, S.

### 3.1.2. Center shift

Center shift is a very important spectral parameter, as its temperature dependence permits determining of the Debye temperature, $T_B$, hence gives some insight into a lattice dynamics. Corresponding behaviors determined based on the analysis A are presented in Figs. 5a and 5b for CS1, CS2 and CS3, whereas Figs. 5c and 5d show a temperature behavior of CS12 and CS, as deduced from the spectra analyzed in terms of procedure B.

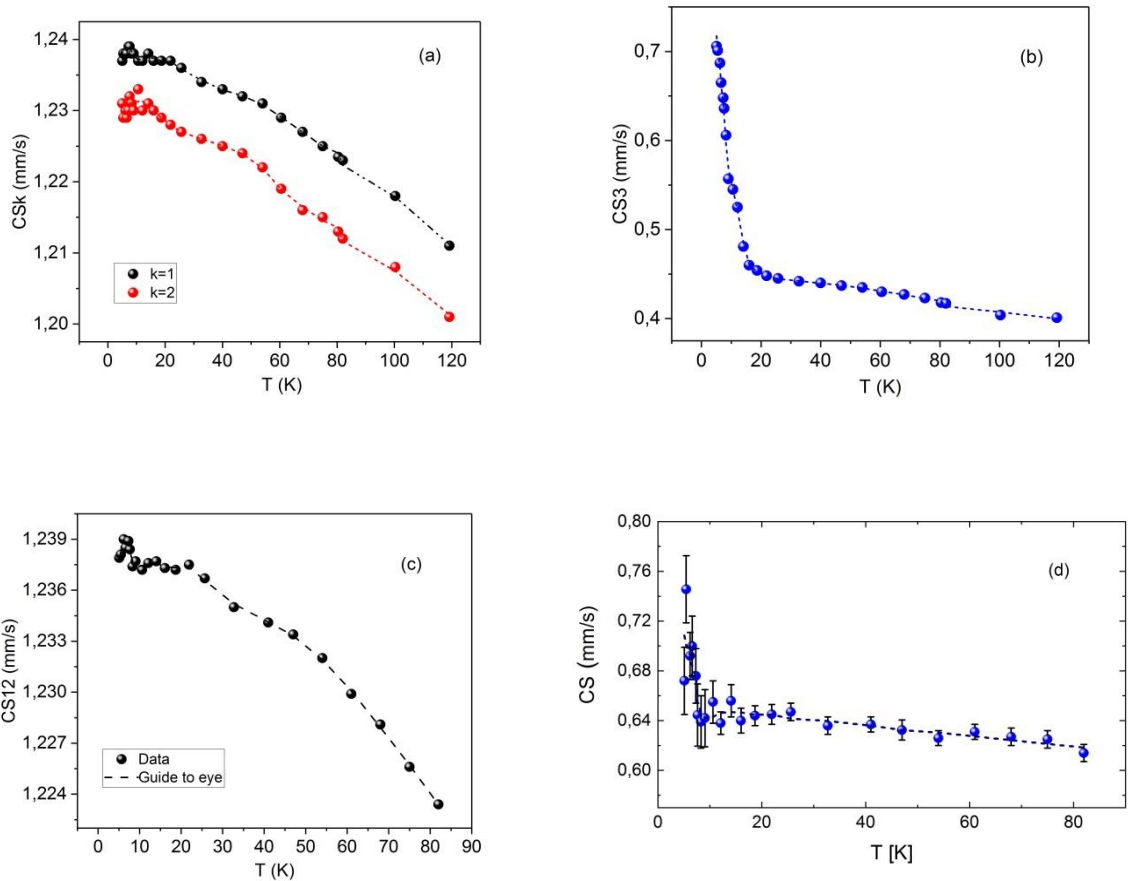

Fig. 5 Temperature dependence of the center shift, $CSk$ (k=1,2, 3) as found using procedure A (a) and (b), and that of $CS12$ and $CS$ using procedure B (c) and (d). Lines stand to guide the eye. Note a strong anomaly for $CS3$ and $CS$ at T < ~20K.

As can be easily noticed in all cases, there are anomalies in the low temperature range, and the strongest ones exist in *CS3* and *CS* i.e. the components related to the minor (ferric) phase. An analysis of the center shift data in terms of the Debye model is given and discussed below.

### 3.1.3. Line width

Figures 6a and 6b illustrate how the line width of the three sub spectra obtained with the two fitting procedures depend on temperature. It is evident that the line width of the sub spectra associated with the major (ferrous) phase are temperature independent while the line width of the minor component exhibit a steep increase below ~25 K. Also, according to both figures, the line width of the D2 doublet is significantly wider than the one of the D1 doublet. This likely indicates that the environment of $Fe^{2+}$ ions is not homogenous and an analysis of the major component in terms of more than two doublets is needed. However, a lack of structure in the shape of lines causes that such analysis would not be unique. A knowledge of the structure of the Fe-gluconate would obviously be helpful in this respect.

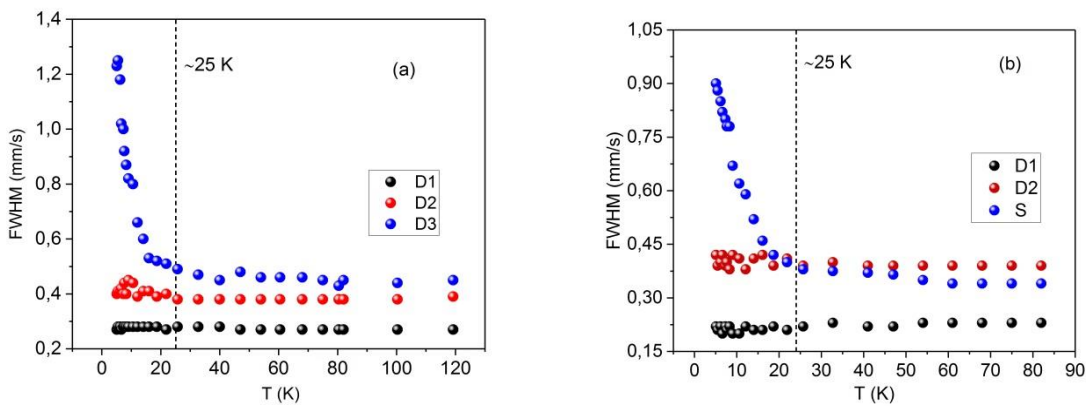

Fig. 6 Full width at half maximum (FWHM) vs. temperature as obtained from the analysis of the spectra in terms of: (a) 3 doublets (D1, D2, D3) and (b) 2 doublets (D1, D2) and 1 sextet (Z1). Note a strong broadening of the line associated with the minor component at *T* below ~25 K.

### 3.1.4. Spectral area

The spectral area is also a useful spectral parameter as it is related to the recoil-free parameter, hence to the lattice vibrations. However, it is the most "difficult" spectral parameter as it is very sensitive to the geometry of experiment and also to the electronic performance and stability of spectrometer. In addition, to get correct information on the spectral area one should analyze spectra using an integral transmission method [14]. In the present case we applied such method with the procedure B. Consequently, only spectral areas obtained with this procedure are shown (Figs. 7a and 7b) and discussed in this paper.

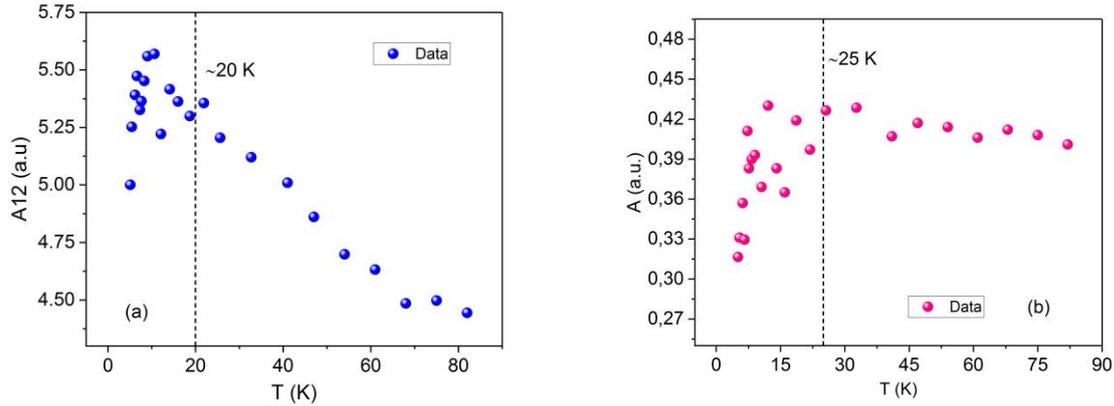

Fig.7 Relative spectral area as obtained with procedure B. Plot (a) shows the spectral area of the major sub spectrum, A12, and plot (b) illustrates the spectral area of the minor sub spectrum, A.

An anomalous behavior can be seen in both plots below ~10-20 K.

**3.2. Debye temperature**

The Debye temperature, $T_D$, can be determined from the Mössbauer spectra in two ways viz. from a temperature dependence of (1) the center shift, $CS(T)$ or (2) the recoil-free fraction, $f(T)$.

**3.2.1. Center shift**

$CS(T)$ can be within the Debye model approximation expressed by the following formula [15]:

$$CS(T) = IS(0) - \frac{3k_B T}{2mc}\left(\frac{3T_D}{8T} + 3\left(\frac{T}{T_D}\right)^3 \int_0^{T_D/T} \frac{x^3}{e^x - 1}dx\right) \quad (1)$$

The first term represents the isomer shift, that hardly depends on *T, m* stays for the mass of the $^{57}$Fe atom, $k_B$ is the Boltzmann constant, *c* is the speed of light, and $x = \hbar\omega/kT$ ($\omega$ being frequency of vibrations). Fitting experimental data to eq.(1) yields the value of $T_D$.

The figures Fig. 8 and Fig.9 illustrate corresponding data and their fits to eq. (1) together with the values of the Debye temperature obtained in high (regular) and low (anomalous) temperature ranges.

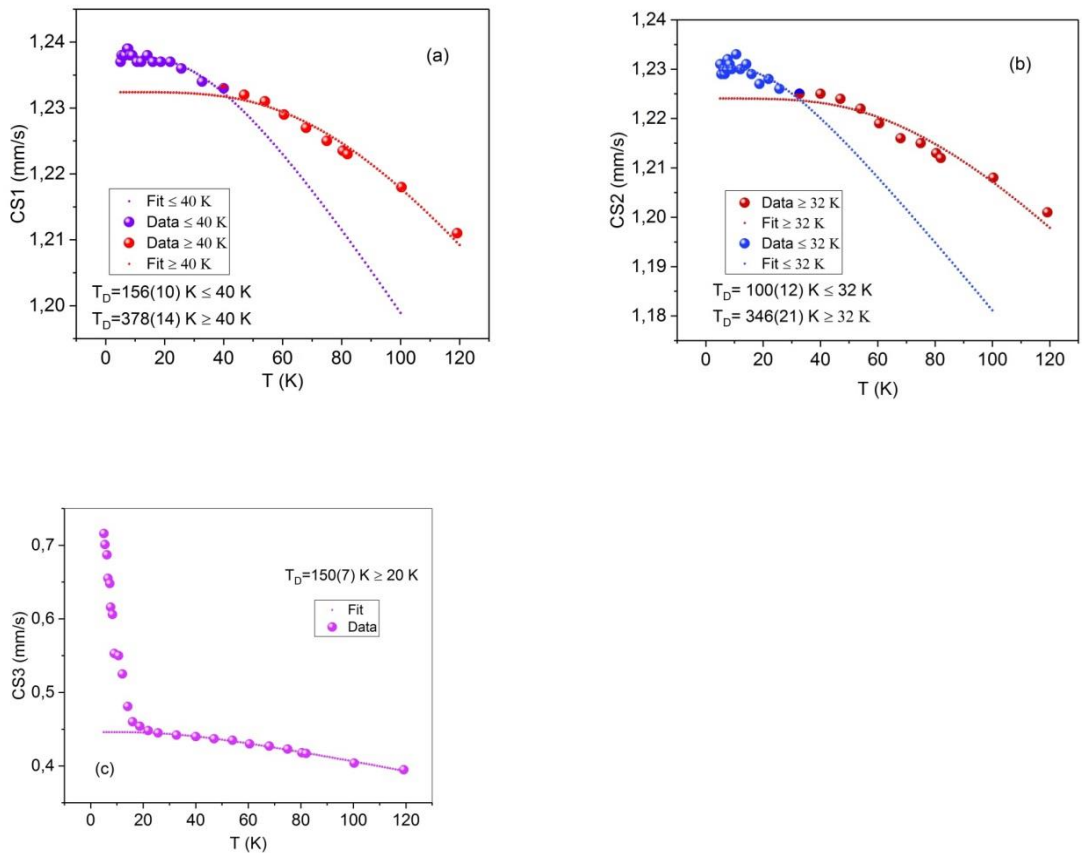

Fig. 8 Temperature dependence of the center shift of the component: (a) D1, *CS1*; (b) D2, *CS2*, and (c) D3, *CS3*. The lines stand for the best fit of the data to eq. (1). Corresponding values of the Debye temperature, $T_D$, derived therefrom are indicated.

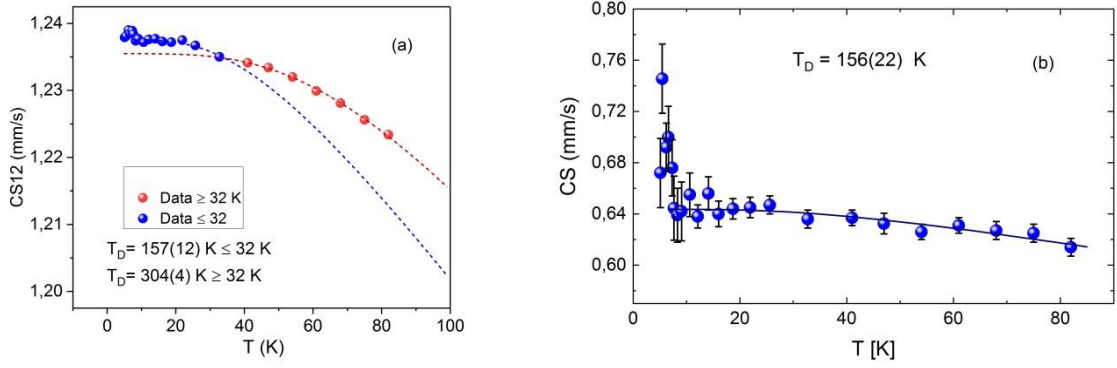

Fig. 9 Temperature dependence of the center shift of the components: (a) D1 and D2, *CS12* and (b) S. The lines stand for the best fit of the data to eq. (1) for T≤ 32 K and for T ≥ 32K. Corresponding values of the Debye temperature, $T_D$, derived therefrom are shown, too.

One can notice that the $T_D$-values characteristic of the low-temperature (anomalous) range of the ferrous phase (major) are by a factor of ~2 lower than those derived from the high-temperature (regular) range. In the ferric phase (minor) the value of the Debye temperature in the non-magnetic phase is, in turn, by a factor of ~2 smaller than the corresponding value found for the ferrous component in the similar temperature range. Why it is so remains unclear.

### 3.2.2. Recoil-free fraction

Temperature dependence of the recoil-free fraction, *f(T)*, is, within the Debye model, given by the following equation [16]:

$$f = \exp\left[\frac{-6E_R}{k_B T_D}\left\{\frac{1}{4} + \left(\frac{T}{T_D}\right)^2 \int_0^{T_D/T} \frac{xdx}{e^x - 1}\right\}\right] \quad (2)$$

Where $E_R$ is the recoil kinetic energy, $k_B$ is Boltzmann constant.

By fitting experimental *f*-values to this equation the value of $T_D$ can be found. In praxis, instead of *f*, whose absolute value is difficult to determine, one uses a relative spectral area, *f'=A(T)/A(T₀)*, which is proportional to *f*. Furthermore, it should be noticed that the values of $T_D$ as determined from *CS(T)* are, in general, different than those found from $f(T)$. This follows from the fact that $CS(T)$ contains information on the average square-velocity of atom vibrations, whereas $f(T)$ is related to the average square-amplitude of such vibrations. Therefore a direct comparison of $T_D$-values obtained from these two spectral parameters is not fully justified.

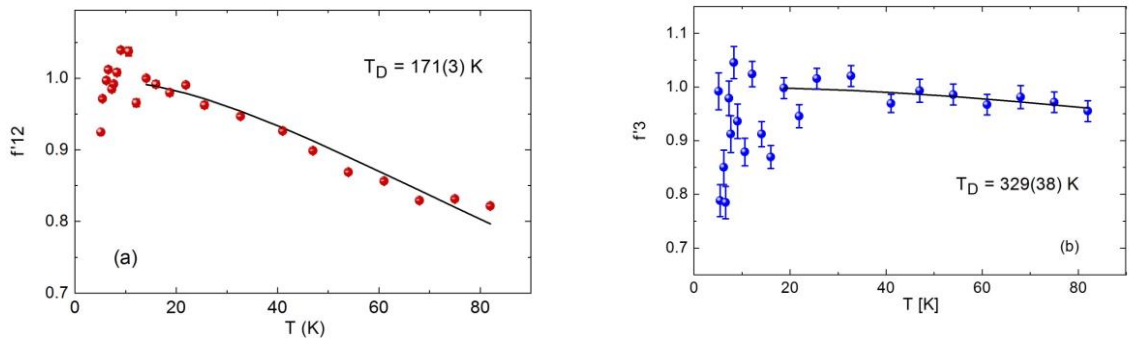

Fig. 10 Temperature dependence of a reduced spectral area for (a) the major component, $f'_{12}$, and (b) for the minor component, $f'_3$. The lines stand for the best fit of the data to eq. (2). Derived therefrom values of the Debye temperature, $T_D$, are shown.

However, we can compare the $T_D$-values obtained with the same method i.e. in the present case the spectral area. As shown in Fig. 10 they are significantly different as determined for the major and for the minor components in the temperature interval where both "phases" are paramagnetic. The difference is twofold in favor of the minor "phase" and this means that the average square-amplitude of ferric ($Fe^{3+}$) ions is shorter than the one of the ferrous ($Fe^{2+}$) ions. Noteworthy, the opposite is true as far as the average square-amplitude of velocity of Fe-ions vibrations is concerned.

### 3.3. Magnetic origin of the line broadening in D3?

A broadening of a line in a Mössbauer spectrum observed on lowering temperature indicates, in general, a transition into a magnetic state. If the magnetism is strong enough, the broadening may result in a split of the line into sextet. In the present experiment the lowest temperature we achieved in Mössbauer measurements was 5 K and the line of the minor sub spectrum, in which the broadening happened, did not change into sextet. Possible reasons for this are a weak magnetism and/or not low enough temperature.

In order to shed more light on the issue, SQUID measurements of the temperature dependence of the magnetic susceptibility $\chi(T)$ along with the magnetization $M$ vs. $H$ data were performed in the temperature range of 2–300 K. Zero-field-cooled (ZFC) and field-cooled (FC) susceptibility curves given in Fig.11a show the characteristic paramagnetic behavior. The same behavior was revealed by AC susceptibility measurements (not shown in Fig.11a for clarity). The $\chi(T)$ curves collapse and can be described with the Curie-Weiss formula. The Curie constant $C=\frac{Ng^2\mu_B^2}{3k_B}S(S+1)$, where $N$ is the Avogadro number, $g$ - Lande factor, $\mu_B$ - Bohr magneton and $S$ is the spin value, obtained by fitting the 20-300 K data is equal to 2.8±0.1 Oe$^{-1}$mol$^{-1}$K. This value is approximately equal to 3 Oe$^{-1}$mol$^{-1}$K expected for the sample containing magnetic centers $S=2$ or close to 3.1 Oe$^{-1}$mol$^{-1}$K in the case of 90% content of the major phase ($S=2$) and 10% of the minor one ($S=5/2$), as concluded from the study above. The paramagnetic Curie-Weiss temperature $\theta$ obtained from the fit equals to -2.6±0.2 K, which points to the antiferromagnetic coupling of the magnetic moments in the sample. This negative interaction, leading to the reduction of the effective magnetic moment $\mu_{eff}=g\sqrt{S(S+1)}=\sqrt{3k_B\chi T/N\mu_B^2}$ is well visible in Fig.11a viz. upon cooling from the room temperature the moment does not change down to about 35K and then decreases, due to the antiferromagnetic coupling setting in at low temperatures. The negative correlations do not however lead to the magnetic order in the whole sample, as no magnetic transition was observed even at the lowest temperatures.

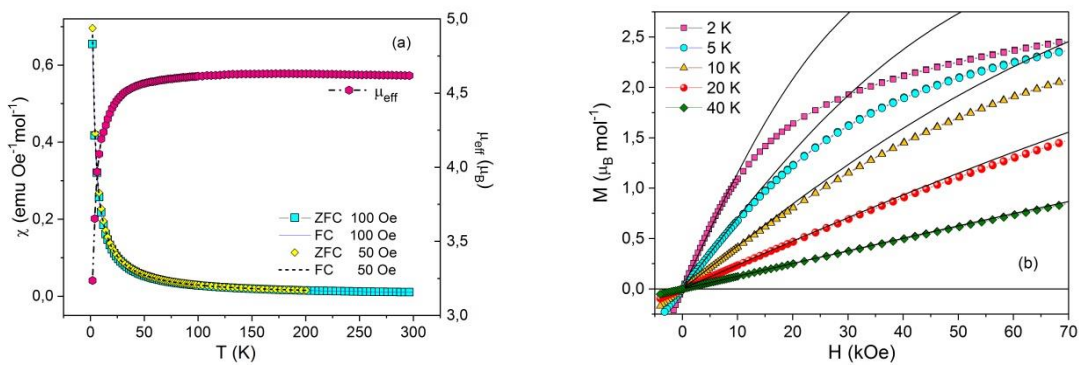

Fig.11 (a) Temperature dependence of magnetic susceptibility measured in $H$=50 Oe and 100 Oe in ZFC and FC modes, as well as the effective magnetic moment expressed per mol of the compound; (b) Magnetization curves measured at several temperatures as compared to the Brillouin function already corrected for the non-zero, negative Weiss temperature.

Magnetization curves recorded at *T* equal 2 K, 5K, 10K, 20K, and 40K (see Fig.11b) differ significantly from the *S*=2 paramagnetic dependences. Experimental *M(H)* data are diminished and this effect gets stronger on lowering the temperature. The thin black lines, linked to experimental data, are the Brillouin curves expected for spin *S*=2, where the negative coupling has been accounted for by putting to the Brillouin function (*T*- θ) instead of *T* [17]. One can see that such simple approach is justifiable only for 40 K and 20 K, while the possible short range magnetic order at *T* < ~20 K needs a more advanced description.

Based on the results of the magnetic measurements the analysis of the spectra in terms of the procedure B was justified. Figure 12 illustrates a temperature dependence of the hyperfine field, *B*, where its increase below ~20 K is evident. The maximum increase is ~7.5 kGs. It should be here added that values of *B* ≠ 0 observed at T ≥~20 K are an artefact caused by the analysis of a doublet in terms of a sextet.

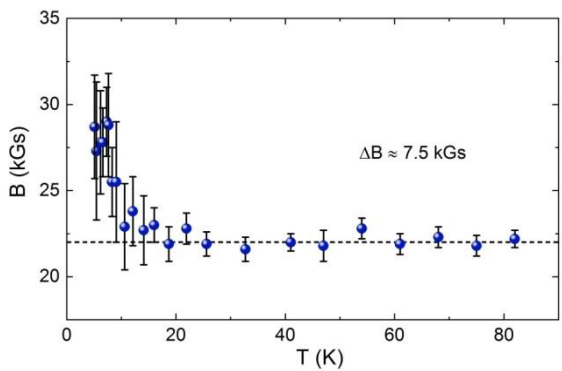

Fig. 12 Temperature dependence of the hyperfine field, *B*. The horizontal line is a reference line as explained in the text.

## 4. Conclusions

Several interesting conclusions can be drawn based on the results obtained in the present study. Namely:

- Iron in the investigated sample of the Fe-gluconate exists in two "phases": a major one with ~90-94% relative contribution at 5K, and a minor one with ~6-10% relative contribution at 5 K.

- Iron in the major "phase" exists as $Fe^{2+}$ ions and as $Fe^{3+}$ in the minor "phase".

- Spectrum associated with the minor phase shows anomalous broadening at temperatures less than ~25 K. This anomaly is reflected not only in the line width and the center shift of the corresponding sub spectrum but also in the center shift of the sub spectra associated with the major "phase".

- Magnetization measurements testify to a magnetic origin of the broadening.

- The maximum value of the hyperfine field derived from the broadening equals to ~7.5 kGs.

- Average square-velocity of Fe ions vibrations in the ferric (minor) phase, is higher than the one in the ferrous (major) phase as the value of the Debye temperature is in the latter is about twofold greater.

- Average square-amplitude of Fe ions vibrations in the ferrous "phase" is shorter than the one in the ferric "phase" as the value of the Debye temperature in the latter is higher by a factor of ~2.

- Lattice dynamics of Fe atoms in the ferrous "phase" seems to be significantly affected by the magnetism of the ferric "phase" as the Debye temperature drops twofold in the temperature range where the magnetism exists.

**Acknowledgements**

This work was financed by the Faculty of Physics and Applied Computer Science AGH UST and ACMIN AGH UST statutory tasks within subsidy of Ministry of Science and Higher Education, Warszawa. Dr. Łukasz Gondek is thanked for recording the XRD pattern.

Table 1 Spectral parameters as obtained with the analysis of the spectra in terms of two doublets D1 and D2 and one sextet. Values of the center shifts (relative to the Rh/Co source at RT) are in mm/s, line width in mm/s, and $V_{zz}$ in

|  | Doublets D1 and D2 | | | | | Sextet (S) | | | | |
| --- | --- | --- | --- | --- | --- | --- | --- | --- | --- | --- |
| T[K] | CS12 | $V_{zz}(1)$ | $V_{zz}(2)$ | G1 | G2 | B [kGs] | CS | $V_{zz}$ | G | A [%] |
| 5.1 | 1.2379(3) | 1.96(1) | 1.80(1) | 0.22(1) | 0.42(1) | 29(3) | 0.67(3) | -0.22(5) | 0.98(8) | 6(1) |
| 5.5 | 1.2381(4) | 1.96(1) | 1.86(1) | 0.21(1) | 0.39(1) | 27(4) | 0.74(3) | -0.27(6) | 0.83(7) | 6(1) |
| 6.2 | 1.2390(5) | 1.96(1) | 1.81(1) | 0.22(1) | 0.40(1) | 28(3) | 0.69(2) | -0.26(5) | 0.70(8) | 6(1) |
| 6.6 | 1.2385(4) | 1.96(1) | 1.86(1) | 0.20(1) | 0.42(1) | 28(3) | 0.71(2) | -0.25(4) | 0.78(8) | 6(1) |
| 7.3 | 1.2389(3) | 1.96(1) | 1.80(1) | 0.22(1) | 0.39(1) | 28(2) | 0.65(2) | -0.22(3) | 0.82(7) | 7(1) |
| 7.65 | 1.2384(4) | 1.96(1) | 1.85(1) | 0.21(1) | 0.40(1) | 29(3) | 0.64(3) | -0.18(4) | 0.83(8) | 7(1) |
| 8.3 | 1.2374(3) | 1.96(1) | 1.79(1) | 0.22(1) | 0.38(1) | 26(2) | 0.64(2) | -0.26(4) | 0.78(5) | 7(1) |
| 9.05 | 1.2377(4) | 1.96(1) | 1.84(1) | 0.20(1) | 0.42(1) | 26(3) | 0.65(2) | -0.23(4) | 0.77(5) | 7(1) |
| 10.6 | 1.2372(3) | 1.96(1) | 1.84(1) | 0.20(1) | 0.43(2) | 23(3) | 0.67(2) | -0.29(5) | 0.62(5) | 6(1) |
| 12.1 | 1.2376(4) | 1.96(1) | 1.78(1) | 0.22(1) | 0.38(1) | 24(2) | 0.64(1) | -0.23(2) | 0.59(4) | 7(1) |
| 14.0 | 1.2377(4) | 1.96(1) | 1.82(1) | 0.21(1) | 0.41(1) | 23(2) | 0.66(1) | -0.26(2) | 0.52(4) | 7(1) |
| 16.1 | 1.2373(4) | 1.96(1) | 1.83(1) | 0.21(1) | 0.42(1) | 23(1) | 0.64(1) | -0.25(2) | 0.42(3) | 6(1) |
| 18.7 | 1.2372(3) | 1.95(1) | 1.77(1) | 0.22(1) | 0.39(1) | 22(1) | 0.64(1) | -0.25(1) | 0.42(3) | 7(1) |
| 21.9 | 1.2375(4) | 1.96(1) | 1.81(1) | 0.21(1) | 0.41(1) | 23(1) | 0.646(8) | -0.25(1) | 0.40(3) | 8(1) |
| 25.7 | 1.2367(4) | 1.96(1) | 1.72(1) | 0.22(1) | 0.38(1) | 21.9(7) | 0.647(7) | -0.26(1) | 0.38(2) | 8(1) |
| 32.7 | 1.2350(4) | 1.95(1) | 1.76(1) | 0.23(1) | 0.38(1) | 21.6(7) | 0.636(7) | -0.25(1) | 0.38(2) | 8(1) |
| 41.0 | 1.2341(4) | 1.95(1) | 1.76(1) | 0.22(1) | 0.39(1) | 22.0(5) | 0.637(6) | -0.25(1) | 0.33(2) | 8(1) |
| 47.0 | 1.2334(4) | 1.95(1) | 1.76(1) | 0.22(1) | 0.39(1) | 21.8(9) | 0.632(8) | -0.26(1) | 0.38(2) | 8(1) |
| 54.0 | 1.2320(4) | 1.95(1) | 1.76(1) | 0.23(1) | 0.39(1) | 22.8(6) | 0.626(6) | -0.24(1) | 0.35(2) | 8(1) |
| 61.0 | 1.2299(4) | 1.95(1) | 1.76(1) | 0.23(1) | 0.39(1) | 21.9(6) | 0.631(6) | -0.25(1) | 0.34(2) | 8(1) |
| 68.0 | 1.2281(4) | 1.95(1) | 1.75(1) | 0.23(1) | 0.39(1) | 22.3(6) | 0.627(7) | -0.25(1) | 0.34(2) | 8(1) |
| 75.0 | 1.2256(4) | 1.95(1) | 1.75(1) | 0.23(1) | 0.39(1) | 21.8(6) | 0.625(7) | -0.25(1) | 0.34(2) | 8(1) |
| 82.0 | 1.2234(4) | 1.95(1) | 1.75(1) | 0.23(1) | 0.39(1) | 22.2(5) | 0.614(7) | -0.23(1) | 0.34(2) | 8(1) |

Table 2 Values of the Debye temperature, $T_D$, as obtained with different analysis procedures and from different spectral parameters.

| Analysis A: Center shift | | |
|---|---|---|
| | T-range [K] | $T_D$ [K] |
| CS1 | 5-40 | 156±0 |
| CS1 | 40-119 | 378±14 |
| CS2 | 5-32 | 100±12 |
| CS2 | 32-119 | 346±21 |
| CS3 | 25-119 | 150±7 |

| Analysis B: Center shift | | |
|---|---|---|
| CS12 | 5-32 | 157±12 |
| CS12 | 32-82 | 304±4 |
| CS | 12-82 | 156±22 |
| Analysis B: Spectral area | | |
| f'12 | 19-82 | 171±3 |
| f'3 | 19-82 | 329±38 |